\documentclass[%
 reprint,
 amsmath,amssymb,
 aps,
]{revtex4-1}

\usepackage{graphicx}
\usepackage{dcolumn}
\usepackage{bm}

\def\Im{\mathop{\mathrm{Im}}}

\def\sub#1{_{\mathrm{#1}}}
\def\up#1{^{\mathrm{#1}}}
\def\Vec#1{\boldsymbol #1}
\def\3He{\mbox{$^3$He}}
\def\4He{\mbox{$^4$He}}
\def\87Rb{\mbox{$^{87}$Rb}}

\begin{document}

\preprint{APS/123-QED}

\title{Vortex Tiling in a Spin-2 Spinor Bose-Einstein Condensate}

\author{Michikazu Kobayashi$^1$, Yuki Kawaguchi$^2$, Masahito Ueda$^{2,3}$}

\affiliation{$^1$Department of Basic Science, University of Tokyo, Komaba 3-8-1, Meguro-ku, Tokyo 153-8902, Japan.}

\affiliation{$^2$Department of Physics, University of Tokyo, Hongo 7-3-1, Bunkyo-ku, Tokyo 113-0033, Japan.}

\affiliation{$^3$ERATO Macroscopic Quantum Control Project, JST, Tokyo 113-8656, Japan}

\date{\today}

\begin{abstract}
We point out that the internal spin symmetry of the order parameter manifests itself at the core of a fractional vortex in real space without spin-orbit coupling.
Such symmetry breaking arises from a topological constraint and the commensurability between spin symmetries of the order parameters inside and outside the core.
Our prediction can be applied to probe the cyclic order parameter in a rotating spin-2 \87Rb condensate as a non-circular vortex core in a biaxial nematic state.
\end{abstract}

\maketitle
Quantized vortices are topological defects of the superfluid order parameter giving rise to the quantized circulation.
In systems with internal degrees of freedom such as spinor Bose-Einstein condensates (BECs) \cite{Ueda}, superfluid \3He \cite{Lounasmaa}, $p$-wave \cite{Jang} and $d$-wave superconductors \cite{Tsuei}, there are various kinds of vortices classified by the topological structures of the order-parameter manifolds.
In particular, the order parameter having discrete symmetry can accommodate a fractional vortex with the circulation that is a fraction of what a vortex in a scalar BEC has.
Examples are half-quantized vortices appearing in a spin-1 polar BEC \cite{Ruostekoski} and the superfluid \3He-A phase \cite{Lounasmaa,Volovik}.
Non-Abelian properties of fractional vortices have also been discussed \cite{Makela,Poenaru,Ivanov}.
Yet another important feature of spinor BECs is that the order parameter at the vortex core may have symmetry different from that of the surrounding.
In such a situation, the following vortex tiling problem arises: what is the core state that is connected smoothly from its surrounding?
For example, the core state of a half-quantized vortex (spin vortex) in a spin-1 polar (ferromagnetic) BEC is ferromagnetic (polar) \cite{Ruostekoski}; the vortex core in the superfluid \3He-B phase at low temperatures and pressures is filled with a planar-like order parameter \cite{Thuneberg,Korhonen,Volovik}; the vortex core of a $d$-wave superconductors is filled with an $s$-wave paring state \cite{Heeb}.

\begin{figure}[tbh]
\centering
\includegraphics[width=0.95\linewidth]{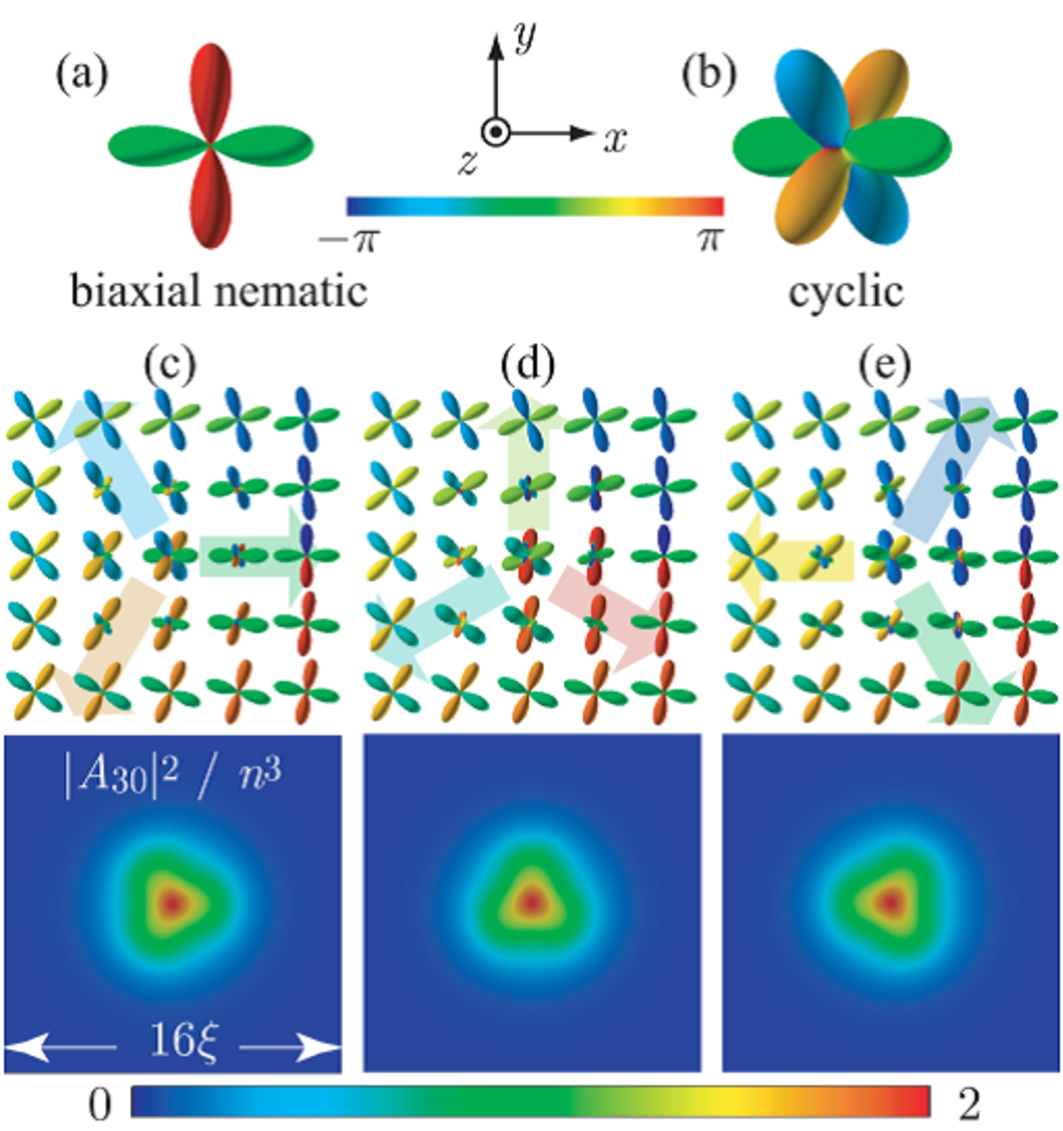}
\caption{\label{fig-tiling} (color) (a) and (b) Plots of $|\Phi(\theta,\phi)|^2$ for the biaxial nematic state $\Psi\sub{B}$ and the cyclic state $\Psi\sub{C}$ defined in TABLE \ref{table-phase}, where $\Phi(\theta,\phi) \equiv \sum_{m=-2}^{2} Y_{2m}(\theta,\phi) \psi_m$. Colors represent the phase of $\Phi(\theta,\phi)$. (c)-(e) Distributions of $\Phi(\theta, \phi)$ (top) and $|A_{30}|^2$ (bottom) for the 1/2--1/4 vortex in the biaxial nematic phase with the cyclic core, where 1/2 and 1/4 refer to the $\pi$ gauge transformation and the $\pi/2$ spin rotation around the vortex, respectively, and $A_{30} =  (3\sqrt{6}/2)(\psi_1^2 \psi_{-2} + \psi_{-1}^2 \psi_2) + \psi_0 (\psi_0^2 - 3 \psi_1 \psi_{-1} - 6 \psi_2 \psi_{-2})$ is the spin-singlet trio amplitude. The size of each panel is $16 \times 16$ in units of the healing length $\xi = \hbar/\sqrt{2 M |c_2| n}$. Arrows with colors in top figures in (c)-(e) indicate the smoothly connected directions between the three lobes of the triad at the core and one of the two lobes of the cloverleaves surrounding the core.}
\end{figure}
In this Letter, we point out that a new situation arises in the vortex tiling problem for a spin-2 BEC which supports more than one possible states having discrete spin (or spin-gauge) symmetry.
When the states inside and outside the core have different discrete symmetries, the internal symmetry of the core state manifests itself in real space as a deformation of the vortex core (see Fig. \ref{fig-tiling}).
Although there is no spin-orbit coupling, the topological constraint, i.e., the global configuration of the order parameter couples the internal spin states to the coordinate space.
Under such a condition, the order parameters inside and outside the core commensurably affect each other, leading to the deformation of the core.
It is also known that the vortex core deforms in a superfluid \3He-B phase and $d$-wave superconductors as a result of splitting of the single vortex into several vortices inside the core \cite{Volovik,Thuneberg,Korhonen} and coupling of the paring state to the anisotropic elementary excitation \cite{Heeb}, respectively.
Thus the origin of the deformation in these systems is the local energetics around the core, and quite different from that in the present case where the core deformation is determined by a global spin configuration outside the vortex core.
Figure \ref{fig-tiling} shows an example of a fractional vortex in the biaxial nematic phase having the 4th dihedral symmetry with the cyclic core having the tetrahedral symmetry.
As shown in Fig. \ref{fig-tiling} (c), the vortex core breaks the rotational symmetry and deforms into a triangular shape, providing an excellent diagnostic method to determine the symmetry of the core state by just looking at the deformed shape of the vortex core \cite{crystal}.
Our prediction can be applied to detect the cyclic state in the vortex core of a rotating spin-2 $^{87}$Rb BEC by preparing the initial state to be in the biaxial nematic state.

\begin{table}[htb]
\centering
\caption{\label{table-phase} $|\Vec{F}|^2$, $|A_{20}|^2$, and $|A_{30}|^2$ and the standard order parameters for the ferromagnetic (F), uniaxial nematic (U), biaxial nematic (B), and cyclic (C) phases.}
\begin{tabular}{ccccl} \hline\hline
Phase & $|\Vec{F}|^2$ & $|A_{20}|^2$ & $|A_{30}|^2$ & order parameter\\ \hline
F & $4n^2$ & 0 & 0 & $\Psi\sub{F} \equiv \sqrt{n} (1,0,0,0,0)^T$ \\
U & 0 & $n^2$ & $n^3$ & $\Psi\sub{U} \equiv \sqrt{n} (0,0,1,0,0)^T / \sqrt{2}$ \\
B & 0 & $n^2$ & 0 & $\Psi\sub{B} \equiv \sqrt{n} (1,0,0,0,1)^T / \sqrt{2}$ \\
C & 0 & 0 & $2 n^3$ & $\Psi\sub{C} \equiv \sqrt{n} (1,0,0,\sqrt{2},0)^T / \sqrt{3}$ \\ \hline
\end{tabular}
\end{table}
We consider a BEC of spin-2 atoms with mass $M$ whose mean-field energy functional is given by \cite{Koashi,Ciobanu}
\begin{align}
H = \int d \Vec{x} \: \left[ h_0 + \frac{c_0}{2} n^2 + \frac{c_1}{2} |\Vec{F}|^2 + \frac{c_2}{2} |A_{20}|^2 \right], \label{eq-Hamiltonian}
\end{align}
where $h_0 = (\hbar^2 / 2M) \sum_{m=-2}^2 |\nabla \psi_m|^2$ is the kinetic energy, $\psi_m$ is the order parameter of a magnetic sublevel $m = 0, \pm 1, \pm 2$ at position $\Vec{r}$; $n = \sum_{m = -2}^{2} |\psi_m|^2$, $\Vec{F} = \sum_{m, m^\prime = -2}^{2} \psi_m^\ast \Vec{f}_{m, m^\prime} \psi_{m^\prime}$, and $A_{20} = \sum_{m=-2}^2 \psi_m \psi_{-m}$ are the total number density, the spin density, and the spin-singlet pair amplitude, respectively, with $\Vec{f}_{m, m^\prime}$ being a vector of spin-2 matrices.
The ground state of Hamiltonian \eqref{eq-Hamiltonian} is (i) ferromagnetic for $c_1 < 0$ and $c_2 > 4 c_1$, (ii) uniaxial nematic or biaxial nematic for $c_2 < 0$ and $c_2 < 4 c_1$, and (iii) cyclic for $c_1 >0$ and $c_2 > 0$.
Each state can be characterized by $\Vec{F}$, $A_{20}$, and the spin-singlet trio amplitude $A_{30} = (3\sqrt{6}/2)(\psi_1^2 \psi_{-2} + \psi_{-1}^2 \psi_2) + \psi_0 (\psi_0^2 - 3 \psi_1 \psi_{-1} - 6 \psi_2 \psi_{-2})$ \cite{Koashi}.
In Table \ref{table-phase}, we list these values and the standard order parameter in each state.
In the mean-field theory, $\Psi\sub{B}$ and $\Psi\sub{U}$ are degenerate at zero magnetic field; however, zero-point fluctuations lift this degeneracy \cite{Song}.
The quadratic Zeeman effect also lifts this degeneracy, stabilizing $\Psi\sub{B}$ for the $F=2$ \87Rb BEC at a magnetic field above a few mG, which is consistent with experiments \cite{Schmaljohann,Chang,Kuwamoto,Widera}.

The order parameter can be expressed in terms of the spherical harmonics: $\Phi(\theta,\phi) = \sum_{m=-2}^{2} Y_{2m}(\theta,\phi) \psi_m$, where $\theta$ and $\phi$ are the polar and azimuthal angles in spin space, respectively.
Figures \ref{fig-tiling} (a) and (b) show $\Phi(\theta,\phi)$ for the biaxial nematic state $\Psi\sub{B}$ and the cyclic state $\Psi\sub{C}$.
The profiles of $\Phi(\theta,\phi)$ for the biaxial nematic and cyclic states features cloverleaf and triad, reflecting the 4th dihedral and tetrahedral symmetries, respectively.
In addition to the trivial $2 \pi$ gauge transformation and the $2 \pi$ rotation in spin space, $\Psi\sub{B}$ remains invariant under the combined operations of the $\pi$ gauge transformation and the $\pi/2$ spin rotation around the $z$ axis, namely $e^{i \pi} e^{- i f^z \pi / 2} \Psi\sub{B} = \Psi\sub{B}$.
This discrete symmetry implies a fractional vortex around which the overall gauge changes from $0$ to $\pi$ and the cloverleaf rotates by $\pi/2$.
We refer to such a vortex as the 1/2--1/4 vortex, where 1/2 and 1/4 refer to the $\pi$ gauge transformation and the $\pi/2$ spin rotation around the vortex, respectively \cite{SO3-SU2}.
The order parameter around a straight 1/2--1/4 vortex on the $z$ axis can be expressed in cylindrical coordinates $(r, \varphi, z)$ as
\begin{align}
\Psi\sub{1/2-1/4} \xrightarrow{r \to \infty} e^{i \varphi / 2} e^{- i f^z \varphi / 4} \Psi\sub{B} = \sqrt{\frac{n}{2}} (1, 0, 0, 0, e^{i \varphi} )^T \label{eq-boundary-1/2}.
\end{align}
Among all kinds of vortices, the 1/2--1/4 vortex is the most stable in the biaxial nematic phase \cite{Yip,Makela}.

\begin{figure}[tbh]
\centering
\includegraphics[width=0.95\linewidth]{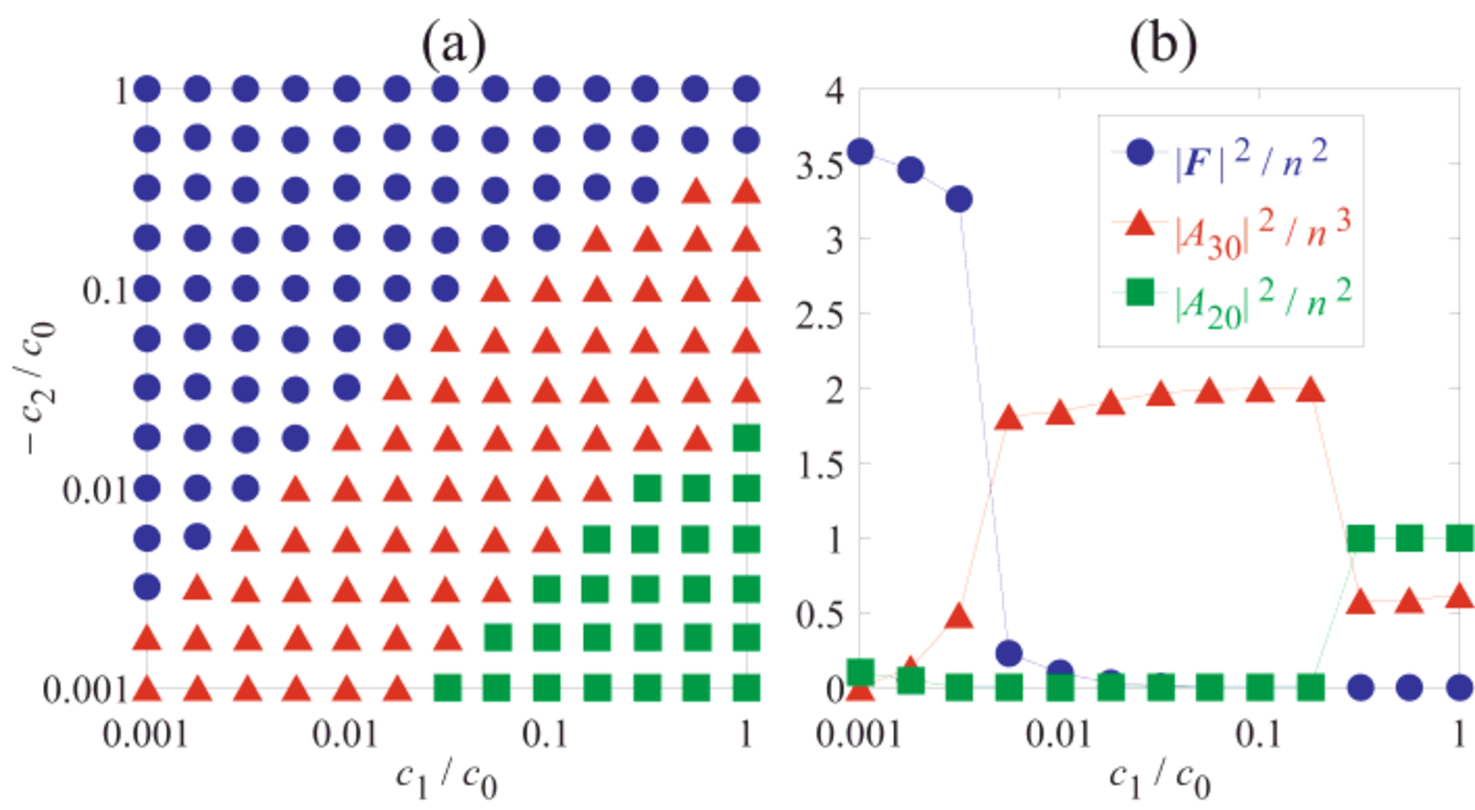}
\caption{\label{fig-vortex-diagram} (color online) (a) Phase diagram of the core state for 1/2--1/4 vortex on the $c_1/c_0$ - $(-c_2/c_0)$, where $\bullet$, {\tiny $\blacksquare$}, and $\blacktriangle$ show the ferromagnetic, uniaxial nematic, and cyclic cores, respectively. (b) $c_1/c_0$ dependence of $|\Vec{F}|^2 / n^2$, $|A_{20}|^2 / n^2$, and $|A_{30}|^2 / n^3$ with $(-c_2/c_0) = 0.01$ at $r = 0$.}
\end{figure}
The core state is determined by the competition among four terms in Eq. \eqref{eq-Hamiltonian} and strongly depends on the interaction parameters $c_{0,1,2}$.
Figure \ref{fig-vortex-diagram} (a) shows a phase diagram of the core state in the $c_1/c_0$ - $(-c_2/c_0)$ plane which is obtained by numerically minimizing the energy functional \eqref{eq-Hamiltonian} with the boundary condition of Eq. \eqref{eq-boundary-1/2}.
Figure \ref{fig-vortex-diagram} (b) shows the $c_1/c_0$ dependence of $|\Vec{F}|^2 / n^2$, $|A_{20}|^2 / n^2$, and $|A_{30}|^2 / n^3$ at $r=0$ along the constant $(-c_2/c_0) =0.01$ line.
These values take almost the same values as shown in Table \ref{table-phase} for each core state and change sharply at the phase boundary.

We now focus on the cyclic core and discuss how it is embedded in the surrounding biaxial nematic state.
In Figs. \ref{fig-tiling} (c)--(e), we show profiles of $\Phi(\theta,\phi)$ and $|A_{30}|^2$ numerically obtained by minimizing the energy functional \eqref{eq-Hamiltonian} with the boundary condition \eqref{eq-boundary-1/2}.
The order parameter is fixed so that the $(0,0,1)$-directions in Figs. \ref{fig-tiling} (a) and (b) coincide with each other, showing the commensurability between the outside cloverleaves and the inside triad.
There are three directions along which one of the three lobes of the triad is smoothly connected with one of two lobes of the cloverleaves surrounding the core.
The triangular core shape is the manifestation of the discrete 3-fold symmetry of the cyclic state.

Under the boundary condition \eqref{eq-boundary-1/2}, the most natural core state with the cyclic order is $\Psi(r=0) = e^{2 i \eta / 3} e^{-i f^z \eta / 3} \Psi\sub{C} = \sqrt{n}(1,0,0,\sqrt{2}e^{i\eta},0)^T/\sqrt{3}$ where $\eta$ is an arbitrary real number and does not affect the energy.
Figures \ref{fig-tiling} (c), (d), and (e) are obtained for $\eta=0, 3\pi/8$, and $3\pi/4$, respectively, and the energies for these three states are degenerate.
Moreover, Figs. \ref{fig-tiling} (c)--(e) show that the direction and the color of the triad at the core changes with $\eta$, resulting in the rotation of the triangular shape of $|A_{30}|^2$.
According to the numerical calculation, the order parameter $\Psi\sub{1/2-1/4}\up{C}$ for the vortex with the cyclic core can be well approximated by
\begin{align}
\Psi\sub{1/2-1/4}\up{C} & = \sqrt{n} (g, \alpha e^{i (\varphi - \eta)}, \beta e^{-i (\varphi - 2 \eta)}, \gamma e^{i \eta}, h e^{i \varphi}),
\label{eq-core-state}
\end{align}
where $g$, $h$, $\alpha$, $\beta$, and $\gamma$ are real functions which satisfy $g = h = 1/ \sqrt{2}$ and $\alpha = \beta = \gamma = 0$ at $r\to\infty$, and $g = 1/\sqrt{3}$, $h = \alpha = \beta = 0$, and $\gamma = \sqrt{2/3}$ at $r = 0$.
From this ansatz, we obtain
\begin{align}
\begin{split}
|\Vec{F}|^2 / n^2 &= (2 h^2 + \alpha^2 - \gamma^2 - 2 g^2)^2 \\
&+ | 2(g \alpha + \gamma h) + \sqrt{6} (\beta \gamma + \alpha \beta e^{i (3 \varphi - 4 \eta)})|^2, \\
|A_{20}|^2 / n^2 &= |2 (g h - \alpha \gamma) +\beta^2 e^{-i (3 \varphi - 4 \eta)}|^2, \\
|A_{30}|^2 / n^3 &= |(3 \sqrt{6} / 2) (g \gamma^2 + \alpha^2 h  e^{i (3 \varphi - 4 \eta)}) \\
&+ \beta^3 e^{-i (3 \varphi - 4 \eta)} - 3 \alpha \beta \gamma - 6 g \beta h|^2,
\end{split}\label{eq-core-value}
\end{align}
and the 3-fold anisotropy and the rotation of the triangular shape of $|A_{30}|^2$ with $\eta$ can be understood by the term $e^{\pm i (3 \varphi - 4 \eta)}$ which rotates the triangular profile by $8\pi/3$ with changing $\eta$ from 0 to $2\pi$.
Since the $\eta$ dependence of the energy functional \eqref{eq-Hamiltonian} disappears by integrating with respect to $\varphi$, $\eta$ is related to the spontaneous breaking of the $U(1)$ symmetry at the core, which implies a Goldstone mode localized at the core.

\begin{figure}[tbh]
\centering
\includegraphics[width=0.95\linewidth]{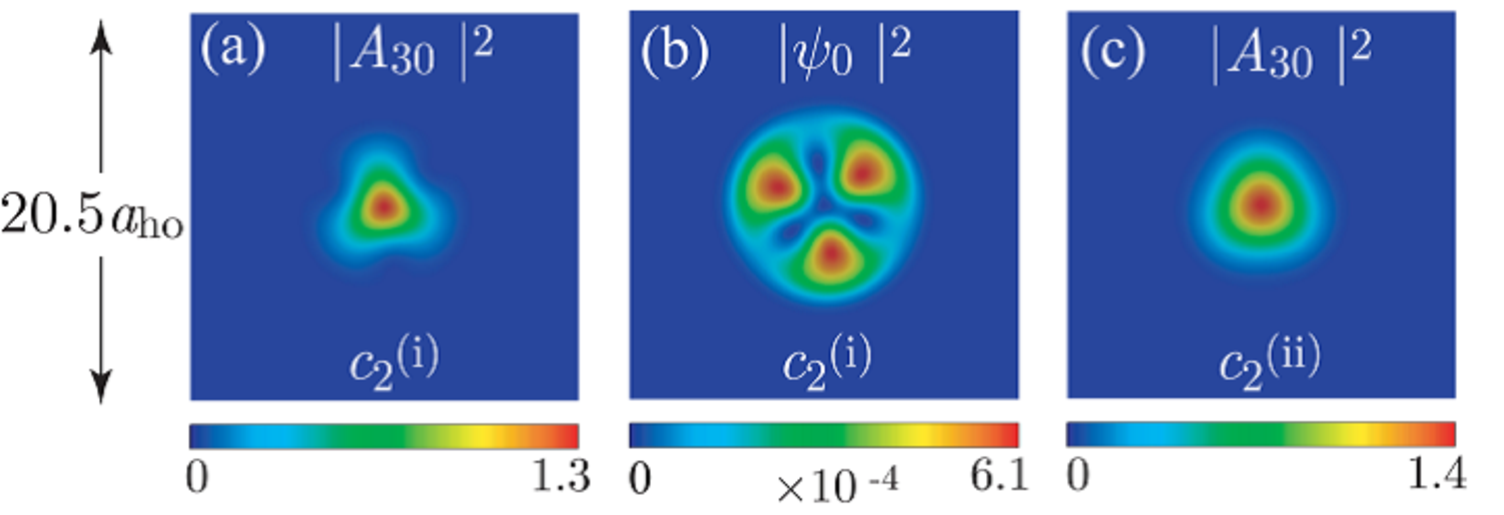}
\caption{\label{fig-rotexp} (color) (a) $|A_{30}|^2$ and (b) $|\psi_0|^2$ of the 1/2--1/4 vortex with the coupling constant $c_2\up{(i)}$, and (c) $|A_{30}|^2$ of the 1/2--1/4 vortex with $c_2\up{(ii)}$. Here $a\sub{ho}$ is the harmonic oscillator length. $|A_{30}|^2$ and $|\psi_0|^2$ are normalized by $n(r=0)^3$ and $n(r=0)$, respectively.}
\end{figure}
We show that the vortices shown in Fig. \ref{fig-tiling} can be generated in the $F=2$ \87Rb BEC under an external rotation.
For an $F=2$ \87Rb BEC, $c_1/c_0 \simeq (8.80 \pm 0.53) \times 10^{-3}$ and $(-c_2/c_0) \simeq (4.71 \pm 5.15) \times 10^{-3}$ were experimentally obtained \cite{Widera}, from which we can expect the 1/2--1/4 vortex with the cyclic core in this system (see Fig. \ref{fig-vortex-diagram}).
It has the nonzero mass circulation $\kappa = \hbar / M \oint d\Vec{l} \cdot \: \sum_m \Im[\psi^\ast_m (\nabla \psi_m)] / n = h / (2 M)$ and can be experimentally generated by an external rotation.
In the presence of a trapping potential $V = M \{\omega_r^2 [x^2 + (1+\varepsilon) y^2] + \omega_z^2 z^2\} n / 2$ with anisotropic parameter $\varepsilon$, the quadratic Zeeman term $q \sum_{m=-2}^2 m^2 |\psi_m|^2$, and external rotation $i \hbar \Omega_z \sum_{m=-2}^2 \psi^\ast_m (x \partial_y - y \partial_x) \psi_m$ along the $z$ axis, we find the stationary state of the order parameter for the $F=2$ \87Rb spinor BEC by minimizing the total energy.
With experimental values of $\omega_r = 141 \times 2\pi$Hz, $\varepsilon=0.05$, $7.0 \times 10^{11}$cm$^{-2}$ for the two-dimensional density at the trap center \cite{Kuwamoto}, and 50mG for the external magnetic field, we perform two-dimensional simulations under $\Omega_z = 0.15 \omega_r$.
The values of $c_{0,1,2}$ are taken from Ref. \cite{Widera}.
For $c_2$, we adopt $(-c_2\up{(i)}/c_0) = 4.71 \times 10^{-3}$ where the ground state is expected to be biaxial nematic.
The stationary state can be obtained by calculating the imaginary-time development of the Gross-Pitaevskii equation derived from Eq. \eqref{eq-Hamiltonian} \cite{Saito}.
As an initial state, we choose stationary $\Psi\sub{B}$ without rotation.
The results are shown in Figs. \ref{fig-rotexp} (a) and (b).
The 1/2--1/4 vortex is stabilized under rotation, and the spontaneous breaking of rotational symmetry also occurs reflecting the symmetry of the core state, which is shown in Fig. \ref{fig-rotexp} (a) for $|A_{30}|^2$ in the 1/2--1/4 vortex.
This symmetry breaking brings about an anisotropic density distribution.
Our predictions can be experimentally tested by the Stern-Gerlach experiment.
In the present simulation, $|\psi_0|^2$ is highly anisotropic as shown in Fig. \ref{fig-rotexp} (b).

We have numerically confirmed that deformation of the vortex core can be observed for the rotational frequency $0.12 \omega_r \lesssim \Omega_z \lesssim 0.17 \omega_r$.
Below $0.12 \omega_r$, no vortices enter the condensate, and above $0.17 \omega_r$, more than one vortex enters the condensate.
The value of $|A_{30}|^2$ (or $|\psi_0|^2$) and its anisotropy decrease with increasing the magnetic field because the cyclic order is weakened by the quadratic Zeeman effect.
Above 70mG, the anisotropy can hardly be seen.
The deformation of the vortex core is hardly affected by the anisotropy of the condensate itself, e.g., by $\varepsilon$ which is necessary to rotate condensate.
Even with $\varepsilon =0.5$, for example, the anisotropic shape of $|\psi_0|^2$ can clearly be seen.

We have also numerically confirmed that the deformation of the vortex core occurs independently of the sign of $c_2$.
Even when we start from the ``false" ground state $\Psi\sub{B}$ with $(-c_2\up{(ii)}/c_0) = -0.44 \times 10^{-3}$, this state can accommodate the ``metastable" 1/2--1/4 vortex, because the values of $c_{1,2}$ are so small that the spin relaxation dynamics from the false to true ground state (cyclic) is much slower than that of the vortex nucleation.
Figure \ref{fig-rotexp} (c) shows the 1/2--1/4 vortex in $\Psi\sub{B}$ with $c_2\up{(ii)}$, and is quite similar to Fig. \ref{fig-rotexp} (a).
These results imply that the nucleated vortex depends only on the initial state regardless of the sign of $c_2$, and the 1/2--1/4 vortex can be experimentally realized using the $F=2$ \87Rb spinor BEC.
In particular, the cyclic state can be realized at the core of the 1/2--1/4 vortex surrounded by the biaxial nematic state.
Since the 1/2--1/4 vortex is non-Abelian \cite{Yip,Makela}, the above method can be used to achieve the first experimental realization of the non-Abelian vortex.

\begin{figure}[tbh]
\centering
\includegraphics[width=0.85\linewidth]{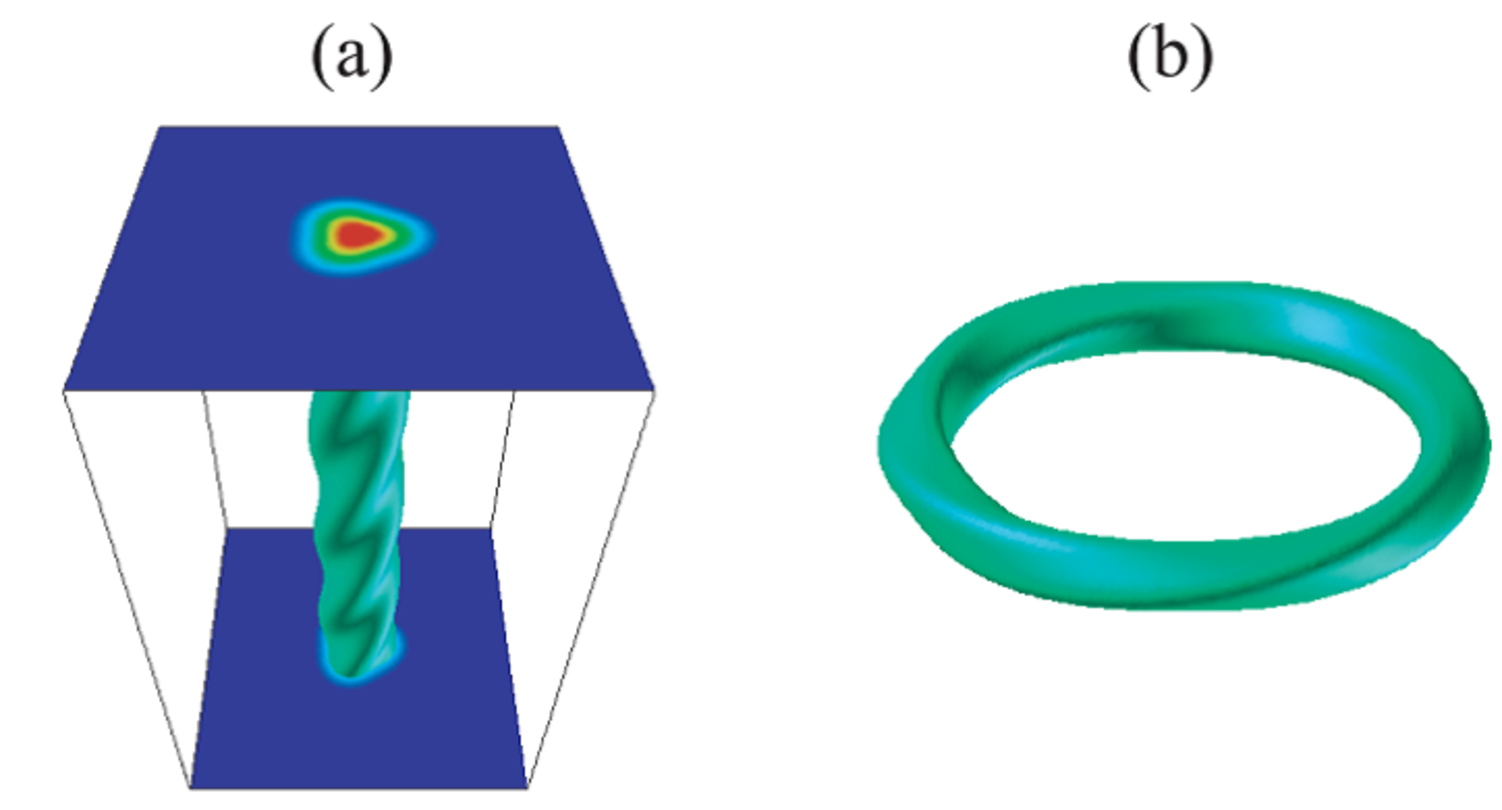}
\caption{\label{fig-twiston} (color online) Isosurface plot of $|A_{30}|^2$ for the states of (a) twisting wave and (b) vorton with $c_1/c_0 = (-c_2/c_0) = 0.01$.}
\end{figure}
The deformation of the vortex core leads to new types of excitations, namely a ``twisting wave" and a ``vorton".
Figures \ref{fig-twiston} (a) and (b) show the numerically obtained twisting wave and stable vorton.
Due to the Goldstone mode with respect to $\eta$ in Eq. \eqref{eq-core-value}, the triangular vortex core can rotate without energy cost.
When we consider a 3-dimensional vortex line, there should be the characteristic wave excitation which corresponds to the gradual change of $\eta$ and the twisting of the triangle shape along the vortex line, and can be regarded as ``twisting wave".
The twisting wave has been studied and experimentally observed for the double-core vortex in \3He-B \cite{Korhonen,Volovik}, and the spin-2 BEC is expected to be another candidate system to investigate the twisting wave, the twisting shape of which can be directly observed.
This twisting wave can create a ``vorton" which is a twisting vortex ring along which $\eta$ changes from 0 to $2 \pi$ and the triangle shape of the core rotates by $8\pi/3$ due to the factor $e^{\pm i (3 \varphi - 4 \eta)}$ in Eq. \eqref{eq-core-value}.
Being different from the usual vortex ring, the vorton also has the mass circulation $2h/(3M)$ along the ring and topologically stable structure.
The vorton has been studied as a stable cosmic string loop \cite{Carter}, and the spin-2 BEC can serve a first theoretical model for the laboratory system to study stable vortons.

In conclusion, we have studied the vortex tiling problem for fractional vortices in a spin-2 spinor BEC.
A 1/2--1/4 vortex with a cyclic core in the biaxial nematic phase should satisfy the commensurability condition between the 4th dihedral symmetry of the biaxial nematic state and the tetrahedral symmetry of the cyclic state.
The rotational symmetry of the cores is spontaneously broken and its shape becomes triangular, which reflects the 3-fold rotational symmetry of the spin state.

This work was supported by KAKENHI (22740219, 22340114, and 22103005), Global COE Program ``the Physical Sciences Frontier", and the Photon Frontier Network Program, MEXT, Japan.


\begin{thebibliography}{99}

\bibitem{Ueda}
M. Ueda and Y. Kawaguchi, arXiv:1001.2072 (2010).

\bibitem{Lounasmaa}
O. V. Lounasmaa and E. Thuneberg, Proc. Natl. Acad. Sci. USA {\bf 96}, 7760 (1999).

\bibitem{Jang}
J. Jang, {\it et al}., Science {\bf 331}, 186 (2011).

\bibitem{Tsuei}
C. C. Tsuei and J. R. Kirtley, Rev. Mod. Phys. {\bf 72}, 969 (2000).

\bibitem{Ruostekoski}
J. Ruostekoski and J. R. Anglin, Phys. Rev. Lett. {\bf 91}, 190402 (2003).

\bibitem{Volovik}
G. E. Volovik, {\it The Universe in a Helium Droplet} (Oxford University Press, New York, 2003), p. 165.

\bibitem{Makela}
H. M\"akel\"a, Y. Zhang and K.-A. Suominen, J. Phys. A: Math. Gen. {\bf 36}, 8555 (2003);
H. M\"akel\"a, J. Phys. A: Math. Gen. {\bf 39}, 7423 (2006);
G. W. Semenoff and F. Zhou, Phys. Rev. Lett {\bf 98}, 100401 (2007);
M. Kobayashi {\it et al}., Phys. Rev. Lett. {\bf 103}, 115301 (2009).

\bibitem{Poenaru}
V. Poenaru and G. Toulouse, J. Phys. (Paris) {\bf 38}, 887 (1977);
N. D. Mermin, Rev. Mod. Phys. {\bf 51}, 591 (1979);
G. E. Volovik and V. P. Mineev, JETP {\bf 45}, 1186 (1977);

\bibitem{Ivanov}
D. A. Ivanov, Phys. Rev. Lett. {\bf 86}, 268 (2001).

\bibitem{Thuneberg}
E. V. Thuneberg, Phys. Rev. Lett. {\bf 56}, 359 (1986);
M. M. Salomaa and G. E. Volovik, Europhys. Lett. {\bf 2}, 781 (1986).

\bibitem{Korhonen}
Y. Kondo, {\it et al}., Phys. Rev. Lett. {\bf 67}, 81 (1991).

\bibitem{Heeb}
R. Heeb, {\it et al}., Phys. Rev. B {\bf 54}, 9385 (1996) and references therein.

\bibitem{crystal}
A similar commensurability problem arises for disclinations in crystals:
See, P. Chaikin and T. Lubensky, {\it Principles of Condensed Matter Physics} (Cambridge University  Press, Cambridge, England, 1995), p. 495.

\bibitem{Koashi}
M. Koashi and M. Ueda, Phys. Rev. Lett. {\bf 84}, 1066 (2000);
M. Ueda and M. Koashi, Phys. Rev. A {\bf 65}, 063602 (2002).

\bibitem{Ciobanu}
C. V. Ciobanu, S.-K. Yip, and T.-L. Ho, Phys. Rev. A {\bf 61}, 033607 (2000).

\bibitem{Song}
J. L. Song, G. W. Semenoff, and F. Zhou, Phys. Rev. Lett. {\bf 98}, 160408 (2007); 
A. M. Turner, {\it et al}., Phys. Rev. Lett. {\bf 98}, 190404 (2007);
S. Uchino, M. Kobayashi, and M. Ueda, Phys. Rev. A {\bf 81}, 063632 (2010);
S. Uchino, {\it et al}., Phys. Rev. Lett. {\bf 105}, 230406 (2010).

\bibitem{Schmaljohann}
H. Schmaljohann {\it et al}., Phys. Rev. Lett. {\bf 92}, 040402 (2004).

\bibitem{Chang}
M.-S. Chang {\it et al}., Phys. Rev. Lett. {\bf 92}, 140403 (2004).

\bibitem{Kuwamoto}
T. Kuwamoto {\it et al}., Phys. Rev. A {\bf 69}, 063604 (2004);
Tojo {\it et al}., Appl. Phys. B {\bf 93}, 403 (2008).

\bibitem{Widera}
A. Widera {\it et al}., New J. Phys. {\bf 8}, 152 (2006).

\bibitem{SO3-SU2}
Strictly speaking, we should use the notation of the $SU(2)$ spin rotation lifted up from the $SO(3)$ spin rotation to classify the vortices.
The notation using the $SU(2)$ spin rotation enables us to distinguish a vortex and its anti-vortex, which is not essential for our discussion because we fix the spin rotation axis as $+z$ for the definition of the 1/2--1/4 vortex.

\bibitem{Yip}
S. -K. Yip, Phys. Rev. A {\bf 75}, 023625 (2007).

\bibitem{Saito}
H. Saito and M. Ueda, Phys. Rev. A {\bf 72}, 053628 (2005).

\bibitem{Carter}
E. Witten, Nucl. Phys. B {\bf 249}, 557 (1985);
R. L. Davis and E. P. S. Shellard, Nucl. Phys. B {\bf 323}, 209 (1989).

\end{thebibliography}
\end{document}